\documentclass[aps,preprint,a4paper,showpacs,showkeys,superscriptaddress]{revtex4-1}

\usepackage{latexsym}
\usepackage{amsmath,amssymb}
\usepackage{graphicx}
\usepackage{subfigure}
\usepackage{xcolor}
\usepackage{hyperref}     
\hypersetup{colorlinks,%
  citecolor=blue,%
  linkcolor=cyan,%
  pdftex}

\begin{document}

\title{Hawking, Fiducial, and Free-fall Temperature of Black Hole on Gravity's Rainbow}

\author{Yongwan Gim}%
\email[]{yongwan89@sogang.ac.kr}%
\affiliation{Department of Physics, Sogang University, Seoul 121-742,
  Republic of Korea}%

\author{Wontae Kim}%
\email[]{wtkim@sogang.ac.kr}%
\affiliation{Department of Physics, Sogang University, Seoul 121-742,
  Republic of Korea}%

\date{\today}

\begin{abstract}
On gravity's rainbow, the energy of test particles deforms the geometry of a black hole in such a way that
the corresponding Hawking temperature is expected to be modified. It means that
the fiducial and free-fall temperatures on the black hole background should also be modified according to
deformation of the geometry.
In this work, the probing energy of test particles is
assumed as the average energy of the Hawking particle in order to study the particle back reaction of the geometry
by using the advantage of gravity's rainbow.
We shall obtain the modified fiducial and free-fall temperatures, respectively.
The behaviors of these two temperatures on the horizon tell us that black hole complementarity is still
well-defined on gravity's rainbow.
\end{abstract}



\maketitle

\section{Introduction}
\label{sec:intro}

There has been much attention to modified dispersion relations in the regime of gravity's rainbow
from the semi-classical point of view of loop quantum gravity \cite{Gambini:1998it, Alfaro:2001rb, Sahlmann:2002qk, Smolin:2002sz, Smolin:2005cz}.
Such modifications could be found in threshold anomalies
in ultra high cosmic rays
and Tev photons \cite{AmelinoCamelia:1997gz, AmelinoCamelia:1997jx, Colladay:1998fq, Coleman:1998ti, AmelinoCamelia:1999wk,AmelinoCamelia:2000zs, Jacobson:2001tu, Jacobson:2003bn}; however,
they are still not established. Moreover,
threshold anomalies are not a generic feature of the modified dispersion relation but
they are only predicted by modified dispersion relation scenarios with a
preferred reference frame \cite{AmelinoCamelia:2002dx}.
In fact, the modified dispersion relations were based on the doubly special relativity
\cite{AmelinoCamelia:2000ge, AmelinoCamelia:2000mn, AmelinoCamelia:2003uc, AmelinoCamelia:2003ex, Magueijo:2001cr, Magueijo:2002am},
which is an extended version of Einstein's special relativity
in the sense that both the Plank length and speed of light should be required to be invariant in any inertial frames.
In connection with this issue, it was claimed that a nonlinear Lorentz transformation in
the momentum space is needed to keep the double invariant constants.
Subsequently,  Magueijo and Smolin  \cite{Magueijo:2002xx}
proposed that the spacetime background felt by a test particle depends on
its energy such that the energy of the test particle deforms the background geometry and
eventually gives modified dispersion relations.
In particular, according to the modified dispersion relations with the generalized uncertainty principle,
 it was shown that the generalized second law of black hole thermodynamics is valid by modifying a relation between the mass and temperature of the black hole \cite{AmelinoCamelia:2005ik}.
Moreover, it was proposed that the brick wall could be eliminated by choosing appropriate rainbow functions \cite{Garattini:2009nq},
and also claimed that a remnant is formed for all black objects in this theory \cite{Ali:2014zea}.
The gravity's rainbow has been extensively studied in order to explore various aspects for black holes and cosmology
\cite{Galan:2004st,  Hackett:2005mb, Aloisio:2005qt, Galan:2006by,  Amelino-Camelia:2013wha, Barrow:2013gia, Ling:2008sy,  Garattini:2011hy,  Garattini:2011fs,  Ali:2014xqa, Garattini:2012ec,  Garattini:2012ca,  Garattini:2014rwa,  Gim:2014ira}.

One of the most important ingredients in thermodynamic analysis of black hole system is to define temperatures consistently
\cite{Hawking:1974sw}.
The Hawking temperature could  be defined by the surface gravity $\kappa(E)$
from the metric of black holes on gravity's rainbow \cite{Ling:2005bp}, and thus
the metric would naturally depend on the energy $E$ of the test particle in terms of the rainbow functions
to modify the dispersion relation.
In the spirit of gravity's rainbow, any probing energy affects the geometry, so
it seems plausible to assert that Hawking radiation deforms the original background geometry.
Hence, if the temperature were regarded as the average energy of test particles on the background of black hole,
then the Hawking temperature should be characterized by the deformed geometry.
Its form would be different from the standard Hawking temperature due to the rainbow effect.
Apart from the Hawking temperature defined at infinity,
one can also consider additional two
different temperatures; the so-called
fiducial temperature and free-fall temperature.
The former is defined in fixed coordinates of an accelerated frame, while
the latter is defined in a free-falling frame.
We expect these two temperatures would be modified like the Hawking temperature according to the rainbow effect,
and thus it would be interesting to study how to obtain the fiducial temperature and free-fall temperature
on the background of black hole on gravity's rainbow.

In section \ref{sec:TH}, we shall elaborate the Hawking temperature on gravity's rainbow.
At first sight, the temperature obtained from the surface gravity in the energy-independent coordinates
seems different from that from the energy-dependent coordinates.
To resolve this conflict, we will find a useful relation between these two temperatures.
Additionally, we shall identify the relation between the energy of test particles and the Hawking temperature. Then,
a proportional constant between the energy and the temperature
will be fixed by using the modified dispersion relation and the uncertainty relation.
In section \ref{sec:Tfid},
our strategy for the calculation of the fiducial temperature is to use the conventional definition, but
written in terms of the energy-dependent coordinates \cite{Magueijo:2002xx}.
For a certain class of rainbow functions, we obtain an explicit fiducial temperature which
becomes the Hawking temperature at infinity.
On the other hand, the free-fall temperature may be derived by employing the Stefan-Boltzmann relation
to relate the energy density with the temperature in a free-falling frame.
Unfortunately, this is not the case, since the conventional Stefan-Boltzmann relation in the proper frame \cite{Tolman:1930zza,Tolman:1930ona} is not appropriate to apply it directly to
quantum black holes, because it was obtained by assuming traceless condition of the energy-momentum tensor.
So the free-fall temperature gives a pathological behavior at the horizon
as discussed in detail in Ref. \cite{Frolov:2011}.
In section \ref{sec:Tff},
we present the generalized Stefan-Boltzmann relation for a non-vanishing trace
in the presence of Hawking radiation. In fact, Hawking radiation is related to
the trace anomaly of the energy-momentum tensors \cite{Christensen:1977jc}.
Then, the free-fall energy density and the free-all temperature will be calculated
on gravity's rainbow.
In section \ref{sec:discussion}, conclusion and discussion will be given.

\section{Hawking temperature}
\label{sec:TH}
Let us start with the modified dispersion relation \cite{Magueijo:2002xx}
 \begin{equation}\label{MDR1} 
 E^2 f(E/E_p)^2-p^2 g(E/E_p)^2=m^2,
 \end{equation}
where $E$,~$p$, $m$ are the energy,  momentum, mass  of a test particle, and
the Planck energy is denoted by $E_p$. We use the natural units as $\hbar=c=k_{\rm B}=1$.
The rainbow functions $f(E/E_p) , ~g(E/E_p)$ satisfy the limits of
$\lim_{E/E_p \rightarrow 0} f(E/E_p) = 1$ and~ $\lim_{E/E_p \rightarrow 0} g(E/E_p) = 1$.
Note that the above modified dispersion relation can be rewritten in the form of
the original dispersion relation such as $\tilde{E}^2 - \tilde{p}^2 = m^2$
by using the transformation,
\begin{equation}
\tilde{E}=f(E)E,~~~  \tilde{p}=g(E)p. \label{rule}
\end{equation}
From now on, we are going to use a two-dimensional metric
in order for exact solubility without losing essential properties of temperatures.
So let us consider the Schwarzschild black hole on gravity's rainbow,
\begin{align}
ds^2 = - F_1(r,E)dt^2+F_2(r,E) dr^2, \label{metric1}    
\end{align}
where the metric functions are $F_1(r,E)=f^{-2}(E)(1-2GM/r)$ and $F_2(r,E)=g^{-2}(E)(1-2GM/r)^{-1}$.
Then the Hawking temperature can be obtained from the surface gravity as \cite{Ling:2005bp}
\begin{align}
T_{\rm H} &= \frac{\kappa_{\rm H}}{2\pi}\label{THform}  \\
&=  \left. \frac{1}{2 \pi} \sqrt{-\frac{1}{2}\nabla^\mu \xi^\nu \nabla_\mu \xi_\nu} \right\rvert _{r=r_{\text{H}}} \\
&=\frac{g(E/E_p)}{f(E/E_p)} \frac{1}{{8\pi G M}}, \label{t1}
\end{align}
where $\xi_\mu$ is the time-like Killing vector and
$r_H$ is the event horizon.

On the other hand, the metric \eqref{metric1} can also be written in terms of the energy-dependent coordinates  as \cite{Magueijo:2002xx}
\begin{equation}\label{metric3} 
ds^2=-\left(1-\frac{2\tilde{G}M}{\tilde{r}}\right)d\tilde{t}^2+\frac{1}{1-\frac{2\tilde{G}M}{\tilde{r}}}d\tilde{r}^2,
\end{equation}
which the transformation is implemented by $\tilde{t}(E)=t/ f(E)$, $\tilde{r}(E)=r/g(E)$, and $\tilde{G}(E)=G/g(E)$,
where the tilde variables are energy-dependent.
From the metric \eqref{metric3}, the Hawking temperature can be derived from the definition of the surface gravity
as
\begin{align}\label{TH1} 
\tilde{T}_{\rm H} =  \frac{1}{8\pi \tilde{G} M}.
\end{align}
Note that it can be shown that the temperature \eqref{TH1} is the same as Eq. \eqref{t1}
if the temperature transformation is assumed as
\begin{equation}
\tilde{T}_{\rm H} =  f(E) T_{\rm H}, \label{new}
\end{equation}
which yields the compatible result with Eq. \eqref{t1} after rewriting it in terms of the un-tilde variables,
\begin{align}
T_{\rm H} = \frac{g(E/E_p)}{f(E/E_p)} \frac{1}{{8\pi G M}} \label{TH2}. 
\end{align}

From the transformation \eqref{new} and the first equation in Eq. \eqref{rule}, one can see that the temperature
$T_{\rm H}$ behaves in the same manner as the energy $E$ on gravity's rainbow.
According to this fact, it is reasonable to relate the temperature to the energy.
If the Hawking radiation were regarded as the energy of test particles,
then the framework of gravity's rainbow would provide the particle back reaction of the geometry effectively and,
consequently, modify the standard Hawking temperature.
However, the test particles with different energies would give different geometries.
So let us choose a single representative energy, that is, the average energy of particles.
Thus it will be proportional to the Hawking temperature
of thermal bath based on the Wien's law \cite{Ling:2005bp},
\begin{equation}
E =\alpha T_{\rm H}, \label{hoho}
\end{equation}
where $\alpha$ is a proportional constant.
On general grounds, it seems non-trivial to determine the constant.
Nevertheless, we shall fix the constant for a specific modified dispersion relation such as
\cite{AmelinoCamelia:2008qg, AmelinoCamelia:1996pj}
\begin{equation}\label{MDR2} 
m^2 = E^2 - p^2 + \eta p^2\left(\frac{E}{E_p} \right)^n,
\end{equation}
where  $\eta$ is a positive rainbow parameter and $n$ is a positive integer.
Then the rainbow functions can be read off from Eq. \eqref{MDR2} \cite{Ali:2014xqa},
\begin{equation}\label{rainbowfunc}
f(E/E_p)=1, \quad  g(E/E_p)=\sqrt{1-\eta \left(  \frac{E}{E_p} \right)^n},
\end{equation}
where $n=2$ for simplicity.

Following the argument in Ref. \cite{AmelinoCamelia:2005ik},
the Heisenberg uncertainty relation can be used
to obtain the momentum of the particle as $p= \Delta p \sim  1/(2 G M)$,
 where the position uncertainty of the particle is $\Delta x \sim 2 G M$.
So the energy can be expressed as $E =\sqrt{(1+4 m^2 G^2 M^2)(\eta G+4 G^2 M^2)^{-1}}$ with $G=1/E_p^2$.
From Eq. \eqref{hoho}, the temperature can be identified as
\begin{equation}\label{TH5}
 T_{\rm H}=  \frac{1}{2\alpha G M}\sqrt{\frac{4G M^2+16 m^2 G^3 M^4}{4G M^2+\eta}}.
\end{equation}
Next, using the rainbow functions \eqref{rainbowfunc} and
the energy-temperature relation \eqref{hoho},
the temperature defined by the surface gravity \eqref{TH2} can be fixed as
\begin{equation}\label{TH6} 
T_{\rm H}=  \frac{1}{8\pi G M}\sqrt{\frac{64 \pi^2 G M^2}{64 \pi^2 G M^2+\alpha^2\eta}}.
\end{equation}
 Note that the Hawking temperature \eqref{TH6} defined at infinity by using 
the surface gravity  method was originally
obtained by assuming a massless scalar field \cite{Hawking:1974sw}.
For the massive case \eqref{TH5}, 
we are actually interested in the case 
for the well-defined semiclassical approximations for which $mM \ll 1$.
In fact, the massive modes will propagate near infinity but it will decay
exponentially there.
Furthermore, the constant $\alpha$ should depend on the mass of particle $m$ and the mass of black hole $M$,
such that it cannot be an universal constant any more.
In these respects, it is reasonable to take the massless limit for simplicity in order to compare Eq. \eqref{TH5} to \eqref{TH6}.
Then, the proportional constant $\alpha$ is uniquely fixed as $\alpha=4\pi$.

After all, the temperature \eqref{TH2} can be expressed as
\begin{equation}\label{TH4} 
T_{\rm H}=  \frac{1}{8\pi G M}\sqrt{\frac{4G M^2}{4G M^2+\eta}},
\end{equation}
which respects the well-known Hawking temperature for $\eta \to 0$.
Note that the above result is different from Eq. \eqref{TH2} in general; however,
they are the same if the energy of probing particles is the average energy of Hawking particles.

In contrast to the standard Hawking temperature,
the Hawking temperature \eqref{TH4} on gravity's rainbow is finite when the mass of black hole vanishes thanks to the
rainbow parameter $\eta$ which plays a role of cutoff.
In the subsequent sections, we will investigate the fiducial temperature and the free-fall temperature by using
the energy-temperature relation \eqref{hoho} and Hawking temperature on gravity's rainbow  \eqref{TH4}.
\section{Fiducial temperature}
\label{sec:Tfid}
The fiducial temperature for the fixed observer in an accelerated frame on a black hole
can be expressed in the form of blue-shifted Hawking temperature by using the time dilation of frequency at different places \cite{w}.
So,  from the metric \eqref{metric3}, the fiducial temperature can be written as
\begin{align}
\tilde{T}_{\rm{FID}} = \frac{\tilde{T}_{\rm H}}{\sqrt{-\tilde{g}_{tt}}}. \notag \\
\end{align}
Note that at $\tilde{r} \to \infty$, $\tilde{T}_{\rm{FID}}=\tilde{T}_{\rm H}$, which means that
the fiducial temperature also follows the same transformation rule as
\begin{equation}
\tilde{T}_{\rm{FID}}=f(E){T}_{\rm{FID}} \label{huhu}
\end{equation}
like Eq. \eqref{new}. Along with $\tilde{G} = G/g(E)$ and $\tilde{r} = r/g(E)$,
one can easily obtain the fiducial temperature as
\begin{align}
T_{\rm{FID}} =  \frac{g(E)}{f(E)} \frac{1}{ 8\pi G M\sqrt{1-\frac{2GM}{r}}},
\end{align}
and the specific choice of rainbow functions \eqref{rainbowfunc} gives
\begin{align}
T_{\rm{FID}} =\sqrt{1-\eta \left(  \frac{E}{E_p} \right)^2} \frac{1}{ 8\pi G M\sqrt{1-\frac{2GM}{r}}}. \label{TFID2} 
\end{align}

\begin{figure}[pt]
  \begin{center}
  \subfigure[{$T_{\rm FID}(M)$ with fixed {\it r}}]{
   \includegraphics[width=0.45\textwidth]{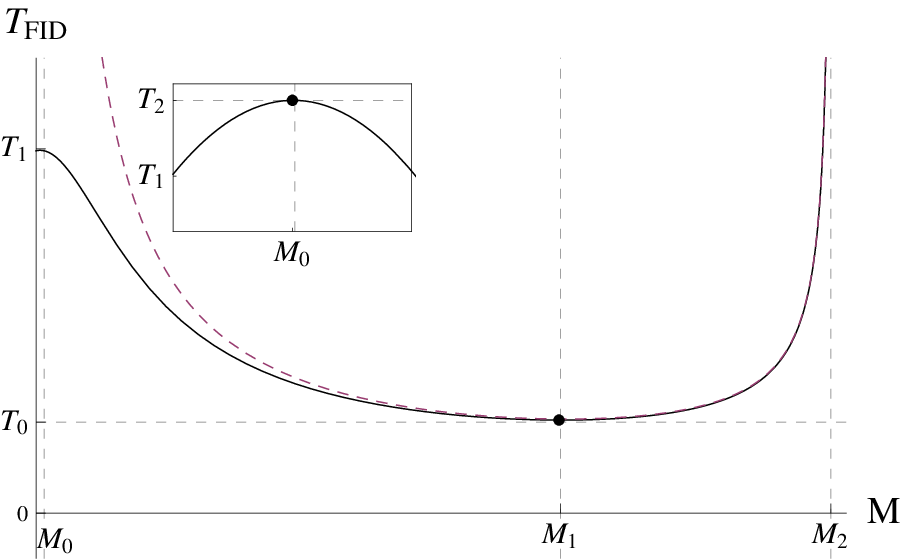}\label{fig:TFIDvsM}}
 \subfigure[{$T_{\rm FID}(r)$ with fixed {\it M}}]{
   \includegraphics[width=0.45\textwidth]{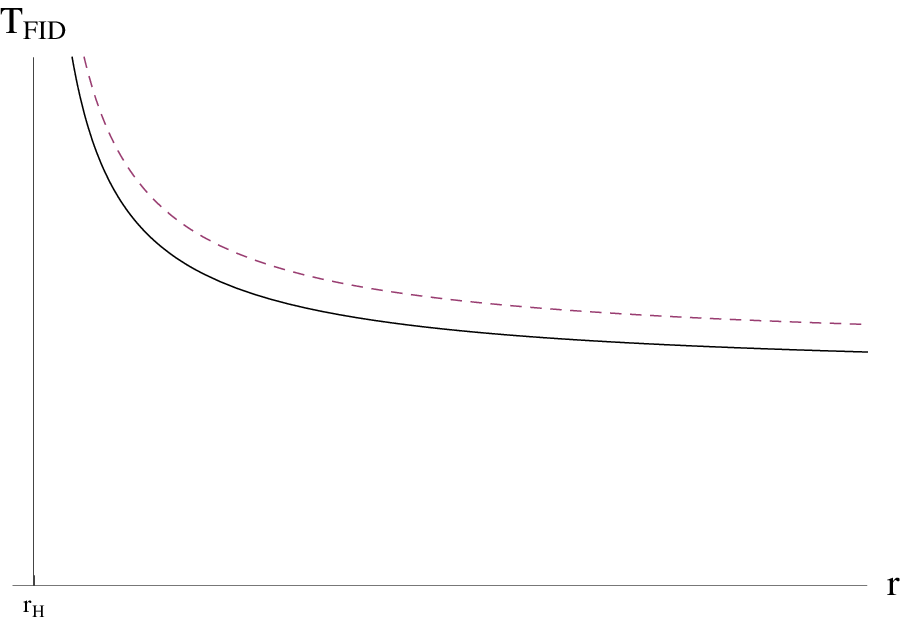}\label{fig:TFIDvsr}}
    \end{center}
  \caption{We set $G=1$, $r=20$, $M_2=10$ in Fig. \ref{fig:TFIDvsM}, and  $M=1$ in Fig. \ref{fig:TFIDvsr}. The
  solid curves are for the fiducial temperatures on gravity's rainbow ($\eta=1$) and
  the dashed curves are for the conventional fiducial temperatures ($\eta =0$).  From the small box in Fig. \ref{fig:TFIDvsM},
  one can see that the fiducial temperature has a finite maximum $T_2$ at $M_0$ and is terminated when $M \to 0$.
  In Fig. \ref{fig:TFIDvsr}, the value of the fiducial temperature is smaller than the conventional one,
  and it is still divergent at the horizon.}
  \label{fig:TvsM}
\end{figure}

In the previous section, we identified the energy of particle as the Hawking temperature.
So, plugging $E= 4\pi T_H$ into the fiducial temperature \eqref{TFID2},
one can explicitly write it as
\begin{align}\label{TFID} 
T_{\rm{FID}} &=\sqrt{1-\eta \left(  \frac{4\pi T_{\rm H}}{E_p} \right)^2} \frac{1}{ 8\pi G M\sqrt{1-\frac{2GM}{r}}}\\
    &= \frac{1}{8\pi G M \sqrt{1-\frac{2G M}{r}}}\sqrt{\frac{4 G M^2}{4G M^2 +\eta}},
\end{align}
where we also used Eq. \eqref{TH4}.
Thus we can show that the fiducial temperature on gravity's rainbow is simply
given as the blue-shifted Hawking temperature {\it i.e.},
$T_{\rm{FID}}=T_{\rm H}/\sqrt{1-2GM/r}$.
In Fig. \ref{fig:TFIDvsM} and \ref{fig:TFIDvsr}, the overall behaviors of the fiducial temperature
on gravity's rainbow are plotted in contrast to the conventional ones.

\section{Free-fall temperature}
\label{sec:Tff}
In this section, we will derive the free-fall temperature defined by a free-falling observer dropped from rest.
First of all,  it is worth to note that the energy density in the free-falling frame should be
rephrased by the temperature through the Stefan-Boltzmann relation which relates the energy density in the
proper frame to the free-fall temperature. In fact,
by using the Stefan-Boltzmann relation, the free-fall temperature
called the Tolman temperature \cite{Tolman:1930zza, Tolman:1930ona}
was obtained assuming the traceless condition of the energy-momentum tensor.
Thus the conventional Stefan-Boltzmann
relation should be generalized in such a way to
incorporate the trace anomaly of energy-momentum tensor \cite{Gim:2015era}, since
Hawking radiation is associated with the trace anomaly \cite{Christensen:1977jc}.

For this purpose, we repeat the calculation along the line of the original work by Tolman \cite{Tolman:1930zza, Tolman:1930ona}
except the traceless condition of energy-momentum tensor.
From the first law of thermodynamics given as
$dU =TdS-pdV$,
where $U$, $T$, $S$, $p$, and $V$ are the thermodynamic internal energy, temperature, entropy, pressure, and
volume in the proper frame, respectively, and $U=\int \rho dV$,
one can get
\begin{equation}\label{first2}
\left.\frac{\partial U}{\partial V} \right\vert_ T  = T \left.\frac{\partial S}{\partial V}\right\vert_T-p.
\end{equation}
By employing the Maxwell relation such as $\partial S/\partial V \vert_T = \partial p/\partial T\vert_V$,
Eq. \eqref{first2} is rewritten as
\begin{equation}
\label{first3}
\rho = T \left.\frac{\partial p}{\partial T}\right\vert_V-p.
\end{equation}
Next the trace of energy-momentum tensor is expressed as
\begin{equation}
-\rho + p = T^\mu_\mu. \label{pp}
\end{equation}
Plugging Eq. \eqref{pp} into Eq. \eqref{first3} along with the property of the temperature-independence of the trace anomaly
 \cite{BoschiFilho:1991xz}, one can get
\begin{equation}
\label{first4}
2\rho = T \left.\frac{\partial \rho}{\partial T}\right\vert_V - T^\mu_\mu,
\end{equation}
which yields the following solutions
\begin{equation}
\rho = \gamma T^2 - \frac{1}{2}T^\mu_\mu,~~~~ p = \gamma T^2 + \frac{1}{2}T^\mu_\mu, \label{sb}
\end{equation}
where the Stefan-Boltzmann constant is chosen as $\gamma= \pi/6$ for a two dimensional massless scalar field \cite{Christensen:1977jc}. This is the generalized Stefan-Boltzmann relation
to incorporate the effect of the trace anomaly. As it should be,
it reproduces the conventional Stefan-Boltzmann relation when the energy-momentum tensor is traceless.

On the other hand, we are now in a position to derive the free-fall energy density and the
pressure. In a static system, the Hawking radiation can be treated as a perfect fluid \cite{Tolman:1930zza},
\begin{equation}\label{Tmunu} 
T^{\mu\nu}=(\rho+p)u^\mu u^\nu +p g^{\mu\nu},
\end{equation}
and from the metric \eqref{metric1} the velocity for the free-falling observer is
solved as
\begin{equation}\label{velocity} 
u^\mu=\frac{dx^\mu}{d\tau} = \left(\frac{1}{ \sqrt{F_1(r,E)}},~~0\right).
\end{equation}
Note that we assumed the Hawking radiation as a perfect fluid
on the static background of black hole. Of course, the background satisfies the equation
of motion from gravity's rainbow without source, whereas the excitations such as Hawking particles
are treated as quantized particles.
Actually, we are interested in the semiclassical limit, such that
the background is indeed classically
vacuum solution whereas the test particles or radiation are quantized by means of the quantum energy-momentum tensor
on this classical background.

Now, the free-fall energy density and the pressure can be related to the quantities defined in the fixed coordinates in terms of
the relations; $\rho=T_{\mu\nu}u^\mu u^\nu$, $p=T_{\mu \nu} n^\mu n^\nu$, where
$n^\mu$ is the spacelike unit normal vector satisfying $n^\mu n_\mu=1$ and $n^\mu u_\mu=0$.
Then the covariant conservation law of the energy-momentum tensor can be written in the form of
\begin{equation}
2 F_1 \partial_r p = -(\rho+p) \partial_r F_1. \label{kk}
\end{equation}
By using the trace equation \eqref{pp},
the solution to the differential equation \eqref{kk} is solved as
\begin{equation}
\rho=\frac{1}{F_1}\left(C_0-F_1 T^\mu_\mu+\frac{1}{2}\int  T^\mu_\mu  dF_1 \right),~~~
p=\frac{1}{F_1}\left(C_0+\frac{1}{2}\int  T^\mu_\mu  dF_1 \right), \label{rhop}
\end{equation}
where $C_0$ is an integration constant.
Plugging the energy density and pressure  \eqref{rhop} into Eq. \eqref{sb},
we can obtain the generalized Tolman temperature as
\begin{equation}\label{Tff1}
T_{\rm FF}=\frac{1}{\sqrt{\gamma F_1}} \sqrt{C_0 - \frac{F_1}{2}T^\mu_\mu+\frac{1}{2}\int  T^\mu_\mu  dF_1 }.
\end{equation}
For the traceless case, the temperature is reduced to the standard Tolman temperature,
$T_{\rm FF}= \sqrt{C_0/(\gamma F_1)}$.
\begin{figure}[pt]
  \begin{center}
      \subfigure[{$T_{\rm FF}(M)$} with fixed $r$]{
  \includegraphics[width=0.45\textwidth]{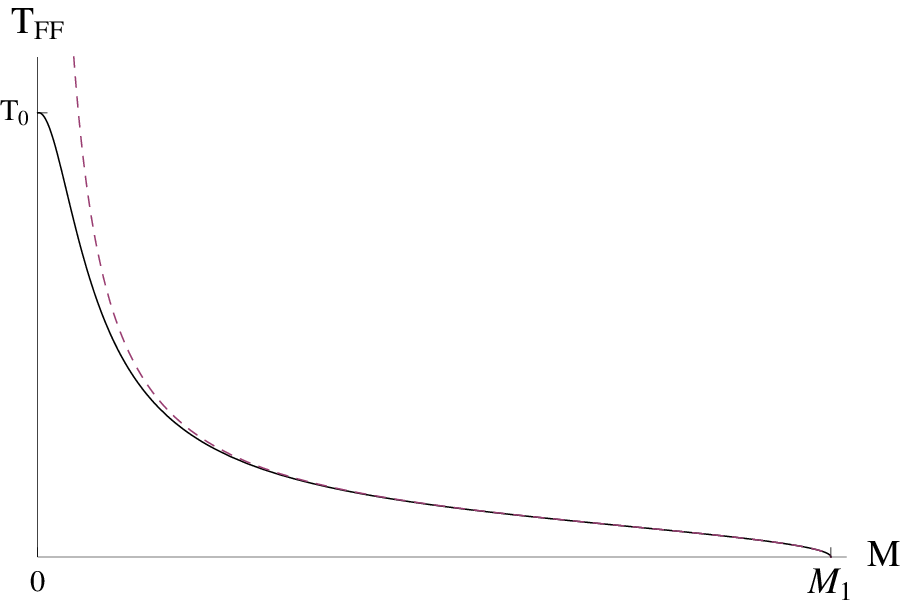}\label{fig:TffvsM}}
\subfigure[{$T_{\rm FF}(r)$} with fixed {\it M}]{
  \includegraphics[width=0.45\textwidth]{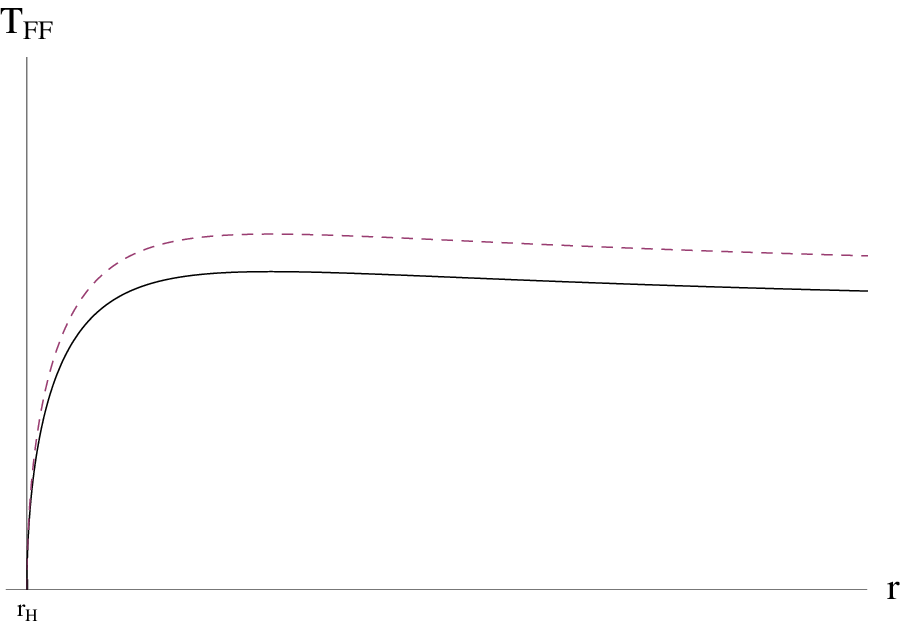}\label{fig:Tffvsr}}
   \end{center}
  \caption{We set $G=1$, $r=20$, $M_1=10$ in Fig. \ref{fig:TffvsM}, and  $M=1$ in Fig. \ref{fig:Tffvsr}.
   The solid curves are for the temperature on gravity's rainbow, which are plotted by setting $\eta=1$ for convenience,
    and the dashed curves are for the conventional temperature without the rainbow effect, simply setting $\eta=0$.}
  \label{fig:Tvsr}
\end{figure}

Let us calculate the free-fall temperature for the two dimensional Schwarzschild black hole
described by the metric \eqref{metric1}.
By using the trace anomaly for a massless scalar field as
$T^\mu_\mu = R/(24\pi)$ \cite{Deser:1976yx,Christensen:1977jc},
the trace anomaly for the metric \eqref{metric1} is obtained as
\begin{equation}
T^\mu_\mu=g(E)^2 \frac{G M}{6\pi r^3}.\label{ano}
\end{equation}
From Eqs. \eqref{Tff1} and \eqref{ano}, the free-fall temperature can be obtained
with the boundary condition of $C_0 = \gamma T_{\rm H}^2$, in which
the standard Hawking temperature is restored at infinity,
\begin{align}\label{Tff}
T_{\rm FF}=\frac{1}{8\pi G M}\sqrt{\frac{4 G M^2}{4G M^2+\eta}}\sqrt{1+\frac{2 G M}{r}+\left(\frac{2 G M}{r}\right)^2-3\left(\frac{2 G M}{r}\right)^3}.
\end{align}
The behaviors of the free-fall temperature are plotted in Fig. \ref{fig:TffvsM} and Fig. \ref{fig:Tffvsr}.
First of all, without the rainbow effect taking $\eta \to 0$,
the temperature is divergent for the massless limit of the black hole due to the rapid evaporation of
black hole but the rainbow parameter cuts off the divergence as seen from Fig. \ref{fig:TffvsM}.
On the other hand, from Fig. \ref{fig:Tffvsr}, one can see that
the temperature becomes the Hawking temperature on gravity's rainbow at infinity, while it
vanishes at the horizon. The radial dependence of the free-fall temperature shows that the rainbow effect
lowers the value of the free-fall temperature.

\section{Conclusion and discussion}
\label{sec:discussion}
The energy of probing particles affects the geometry in the formalism of gravity's rainbow,
which is comparable to take into account the test particle back reaction of the geometry.
It means that the energy of particle modifies the geometry of black hole, so that
the Hawking temperature which is sensitive to the geometry of black hole can also be modified.
In this context, we studied the Hawking, fiducial,
and free-fall temperatures, respectively, in order to
obtain their characteristics in the framework on gravity's rainbow.

For the Hawking temperature, we presented two representations in the fashion of the energy-independent
and -dependent coordinates, and found that the relation to connect these representations
follows $\tilde{T} = f T$ like the energy transformation.
Moreover, the Hawking radiation was identified with using the energy of test particles in order to
investigate the impact on the geometry in the presence of radiation. We found that
the energy-temperature relation was specified as $E =\alpha T_{\rm H}$, where $\alpha=4\pi$ in our choice of
rainbow functions. It is interesting to note that the proportional constant
was given by the irrational number, which is contrast to the common case given as multiple degrees of
half-integer. It means that the particle energy should be non-trivially related to the thermal temperature.
Next, the fiducial temperature was defined by using the energy-dependent coordinates,
and then it was rewritten in terms of the energy-independent coordinates.
Using the energy-temperature relation, we found
that the fiducial temperature \eqref{TFID} takes the blue-shifted Hawking
temperature that is still divergent at the horizon, while it reproduces the Hawking temperature at infinity.
For the free-fall temperature,
we extended the conventional Stefan-Boltzmann relation to the case of the non-vanishing trace of energy-momentum tensor
in order to take into account the trace anomaly related to Hawking radiation.
Consequently, the free-fall temperature is finite everywhere without the blueshift, especially vanishing at the horizon.

In connection with the last statement,
one might want to find a different reason why the free-fall temperature \eqref{Tff} vanishes at the horizon, whereas
the fiducial temperature \eqref{TFID} is divergent there. This fact can also be seen from
the Unruh effect \cite{Unruh:1976db} for the large black hole. Very near the horizon, the metric \eqref{metric3}
can be written as the Rindler metric, so that the acceleration of the fixed observer
is proportional to the temperature as $\tilde{T}_{\rm U}=\tilde{a}/2\pi$. By recovering it in the energy-independent coordinates,
the Unruh temperature on gravity's rainbow can be written as $T_{\rm U}=(g/f) GM/(2\pi r^2  \sqrt{1-2GM/r})$.
From the choice of rainbow functions such as Eq. \eqref{rainbowfunc} with Eq. \eqref{hoho},
we can see that it should be divergent at the horizon,
which is coincident with the present result for the fiducial observer.
As a corollary in the local inertial frame, there does not exist any acceleration  at the horizon for the large black hole,
 so that the Unruh temperature vanishes. It means that
 our free-fall temperature should be zero at the horizon.
Thus the fixed observer and free-fall observer see extremely different degrees of freedom at the horizon,
which means that black hole complementarity \cite{Susskind:1993if, Susskind:1993mu, Stephens:1993an} still holds
for gravity's rainbow.

The final comment is in order.
At first sight, the free-fall temperature generically seems to vanish in freely falling frames
at any distance far from the horizon including at the horizon because
of the equivalence principle. However, this is not the case, since
there exists the energy density which amounts to
the curvature scale of $1/M^2$ even in those frames in which the gravitational acceleration is
locally zero. Thus, the scale of the corresponding temperature is the order of $1/M$ as seen from Eq. \eqref{sb}
rather than zero.
Therefore, the common wisdom is that
the equivalence principle is weakly broken when the Hawking radiation is involved quantum-mechanically.
However, the surprising one is that as shown in Ref. \cite{Singleton:2011vh}
the equivalence principle is restored just at the horizon.
For the large black holes, this fact is compatible with the Unruh effect which was actually defined at the horizon for
those black holes in the Rindler approximation.
In the present calculations, this vanishing result of the free-fall temperature at the horizon was found
from the modification of the Stefan-Boltzmann law by taking into account the trace anomaly.

\acknowledgments
We would like to thank M. Eune for helpful comments.
This work was supported by the National Research Foundation of Korea(NRF) grant funded by the Korea government(MSIP) (2014R1A2A1A11049571).


\bibliographystyle{JHEP}

\bibliography{RGreferences}


%
%

\end{document}